\documentclass[preprint,11pt]{aastex}
\usepackage{amsmath,amssymb,natbib,graphicx,color,ulem,bm}

\bibpunct{(}{)}{;}{a}{}{,}

\newcommand{\s}{\,\mathrm{s}}
\newcommand{\g}{\,\mathrm{g}}
\newcommand{\cm}{\,\mathrm{cm}}

\newcommand{\m}{\,\mathrm{m}}
\newcommand{\mm}{\,\mathrm{mm}}
\newcommand{\km}{\,\mathrm{km}}

\newcommand{\K}{\,\mathrm{K}}
\newcommand{\AU}{\,\mathrm{AU}}

\newcommand{\dyne}{\,\mathrm{dyne}}

\newcommand{\J}{\mathrm{J}}

\def\refnew#1{(\ref{#1})}

\begin{document}

\title{On the mass and origin of Chariklo's rings}
\author{Margaret Pan ({\tt pan@astro.utoronto.ca}), Yanqin Wu}
\affil{Department of Astronomy \& Astrophysics, 50 St. George Street,
University of Toronto, Canada}

\begin{abstract}
  Observations in 2013 and 2014 of the Centaur 10199 Chariklo and its
  ring system consistently indicated that the radial width of the
  inner, more massive ring varies with longitude. That strongly
  suggests that this ring has a finite eccentricity despite the fast
  differential precession that Chariklo's large quadrupole moment
  should induce. If the inferred apse alignment is maintained by the
  ring's self-gravity, as it is for the Uranian rings, we estimate a
  ring mass of a few times $10^{16}$~g and a typical particle size of
  a few meters. These imply a short collisional spreading time of
  $\sim$$10^5$~years, somewhat shorter than the typical Centaur
  dynamical lifetime of a few Myrs and much shorter than the age of
  the solar system. In light of this time constraint, we evaluate
  previously suggested ring formation pathways including collisional
  ejection and satellite disruption.  We also investigate in detail a
  contrasting formation mechanism, the lofting of dust particles off
  Chariklo's surface into orbit via outflows of sublimating CO and/or
  N$_2$ triggered after Chariklo was scattered inward by giant
  planets. This latter scenario predicts that rings should be
  common among 100-km class Centaurs but rare among Kuiper belt
  objects and smaller Centaurs. It also predicts that Centaurs should
  show seasonal variations in cometary activity with activity maxima
  occurring shortly after equinox.
\end{abstract}

\section{Introduction}\label{sec:intro}

Occultation observations of the Centaur 10199 Chariklo on 2013 June 3
revealed two dense narrow rings, the first discovered around a minor
planet \citep{bragaribas14}. This surprising announcement immediately
raised questions about the formation, lifetime, and ubiquity of rings
around small bodies with active dynamical histories: Centaurs are
thought to be Kuiper belt objects (KBOs) scattered into the giant planet
region by planetary encounters, and their typical dynamical lifetime
is $\sim$few~Myr \citep{bailey09}.

The discovery observation resolved the inner and more
massive\footnote{The outer ring, which \citet{bragaribas14} believe
  has about one-tenth the inner ring's mass, was only just resolved by
  the discovery occultation, and its width measurements are much less
  precise.} of the rings and, intriguingly, showed a significant
difference between the ring widths measured during ingress ($7.17\pm
0.14$~km) and egress ($6.16\pm0.11$~km). Further occultations in 2014
again indicated azimuthal variations in the width of the inner ring
\citep{sicardy14}.  If these variations are long-lived, they strongly
suggest an apse-aligned ring with a finite eccentricity spread and,
therefore, a finite overall eccentricity \citep[see, for
  example][]{nicholson78}. This apse-alignment is surprising since
Chariklo's large oblateness
\citep[$\varepsilon=0.213$,][]{bragaribas14} and implied $J_2$ moment
should cause fast differential precession within the ring.

This feature provides us with a convenient way to measure the ring
mass.  Inspired by the work of \citet{goldreich79a,goldreich79} on the
narrow dense Uranian rings, we describe here a simple model of
Chariklo's inner ring in which Chariklo's oblateness and the ring's
self-gravity together maintain the ring's apse alignment. We apply
this model to derive a mass, typical particle size, and collisional
spreading time for the ring (\S \ref{sec:selfgravity}). We discuss our
results' implications for existing dynamical/collisional ring
formation models (\S \ref{sec:formation}), then propose and discuss a
completely different model where the rings result from volatile
outgassing (\S \ref{sec:outgas}).  Finally, we summarize our findings
(\S \ref{sec:summary}).

\section{Ring mass}\label{sec:selfgravity}

\subsection{Setup}\label{sec:setup}

Despite its small size \citep[equatorial radius
$R\sim 145$~km,][]{bragaribas14}, Chariklo appears quite
oblate. Assuming circular rings whose center and orbit normal coincide
with Chariklo's center and axis of symmetry, \citet{bragaribas14} found
an oblateness of $\varepsilon=0.213\pm 0.002$. Under the simplest
assumption of uniform density, Chariklo's lowest order gravitational
moment would be $J_2=\varepsilon(2-\varepsilon)/5\simeq 0.076$.  If
Chariklo has bulk density $\sim$1~g~cm$^{-3}$, the implied differential
precession timescale for a ring with semimajor axis $a\simeq 390$~km
and radial width $\Delta a\simeq 7$~km (see Table
  \ref{table:chariklo}) would be of order
\begin{equation}
\frac{1}{J_2}\frac{a^2}{R^2}\frac{a}{\Delta a}\frac{1}{\Omega}\simeq 17\;\mathrm{months}\;\;\;,
  \label{eq:j2_oom}
\end{equation}
far shorter than any plausible system lifetime.

At the same time, multiple occultations in 2013 and 2014 indicate that
the width of Chariklo's inner, more massive ring varies with longitude
from $\sim$5.5~km to $\sim$7.1~km
\citep{elmoutamid14,sicardy14}. Given the short orbit period
($\sim$16.4~hrs), this strongly suggests an eccentric apse-aligned
ring whose eccentricity varies by at least $\Delta
e\sim$$\delta(\Delta a)/a\simeq(7.1-5.5)/390.6\simeq0.004$ over the
ring's total width.The mean eccentricity $e$ must be at least $\Delta
e$ --- in a similar context, the rings of Uranus have eccentricities 7
to 11 times larger than $\Delta e$ \citep{french91}.

To maintain apse alignment in Chariklo's rings, other forces must
counter the dispersive effects Chariklo's $J_2$. One possibility is
self-gravity within the ring, originally suggested by
\citet{goldreich79,goldreich79a} to explain the apse alignment of the
Uranian rings. To see this qualitatively, we think of the ring as a
collection of non-crossing radially nested ringlets, a reasonable
model for a cold system with frequent collisions; self-gravity
causes outer ringlets to feel an inwards radial force while inner
ringlets feel an outwards radial force. In an eccentric ring that is
narrower at periapse than at apoapse, the self-gravity forces at
periapse dominate over those at other longitudes, so we can
approximate these forces as impulses at periapse. Such forces tend to
enhance the forward precession of the outer ringlets and slow that of
the inner ringlets, cancelling the differential precession due to
Chariklo's quadrupole.

Given the observed system parameters, we can solve for the ring mass
$m$ needed to maintain apse alignment. We give an order of magnitude
estimate here (ahead of an exact solution in \S \ref{sec:exact}) by
dividing a narrow ring into just two apse-aligned ringlets of masses
$m/2$, eccentricities $e$ and $e+\Delta e$, where $0 < \Delta e\ll
e\ll 1$, and semimajor axes $a$ and $a+\Delta a$, where $\Delta a \ll
a$. We find the total gravitational force between the ringlets by
estimating the force at each point and summing over longitude. For a
point on, say, the inner ringlet at true anomaly $f$, the force arises
mostly from a segment of the outer ringlet at the same true anomaly at
a distance $d \sim \Delta a (1-q\cos f)$ away where
\begin{equation}
q = {\frac{d (ae)}{da}} \approx {e + {\frac{a \Delta e}{\Delta a}}}
\sim a {\frac{\Delta e}{\Delta a}}\, .
\label{eq:defineq}
\end{equation} 
Non-crossing orbits in a ring narrower at periapse than apoapse
require $0<q\leq 1$. Since $d\ll a$ we treat the segment on the outer
ringlet as an infinite wire to find
\begin{equation}
  \textrm{self-grav.\;accel.}\sim\frac{Gm}{\pi a\Delta a(1-q\cos f)} +
  O(e^2) \;\;\; .
\end{equation}
This acceleration is mostly radial; the tangential component is
smaller by a factor of order $e$.  Lagrange's equations
\citep{murray99} then give an orbit averaged precession rate of
\begin{equation}
  \left.\frac{d\varpi}{dt}\right|_\mathrm{self}
  \sim \int_\mathrm{ringlet} df\,\frac{\cos f}{\Omega a e}\cdot \frac{Gm}{a\Delta a(1-q\cos f)}
  \sim \frac{Gm}{a (\Delta a)^2\Omega}\frac{\Delta e}{e} \, .
  \end{equation}
  Setting this equal to the reciprocal of the right-hand side of
  equation~\ref{eq:j2_oom} yields
\begin{equation}
  m \sim M\,J_2\frac{e}{\Delta e}\left(\frac{R}{a}\right)^2\left(\frac{\Delta a}{a}\right)^3
  \sim 5\times 10^{15}\;\g\left(\frac{e/\Delta e}{10}\right) \;\;\;.
\end{equation}
where we assume Chariklo is an oblate spheroid with obliquity $0.213$,
constant bulk density $\rho=1$~g~cm$^{-3}$ and total mass $M\simeq
10^{22}$~g. Note that finite $e$ and $\Delta e$ are crucial to the
calculation of $d\varpi/dt|_\mathrm{self}$.

\subsection{Exact solution}\label{sec:exact}

Following the procedure of \citet{goldreich79}, we divide a narrow
apse-aligned eccentric ring into $N$ ringlets and compute the
interaction between ringlets $j,k$. Let ringlet $j$ have semi-major
axis $a_j$, eccentricity $e_j$, and fraction $h_j$ of the total ring
mass $m$. To lowest order in eccentricity, the force from ringlet $k$ on a point on ringlet $j$ is
\begin{equation}
{\bf{F}}_\mathrm{point} = \frac{Gmh_k}{2\pi a_k(1 +e_k\cos f)}\frac{
  \left(\hat{\bf r} -\hat{\bm \theta}\sin f\right)}{(a_k-a_j)(1-q_{jk}\cos  f)}\, ,
\end{equation}
where $\hat{\bf r}$, $\hat{\bm \theta}$ are the radial and
tangential unit vectors and
\begin{equation}
  q_{jk}
  =\left.\frac{d(ae)}{da}\right|_{jk}
  \simeq \frac{a(e_k-e_j)+e(a_k-a_j)}{a_k-a_j} \;\;\; .
\end{equation}
This force induces a precession rate \citep{murray99}
\begin{equation}
\left.\frac{d\omega}{dt}\right|_\mathrm{point} = \frac{1}{\Omega_j
  a_je_j}\left(-\hat{\bf r}\cos f + \hat{\bm \theta}\sin f(2-e_j\cos
  f)\right)\cdot{\bf{F}}_\mathrm{point}\, ,
\end{equation}
so the precession rate of ringlet $j$ due to ringlet $k$'s gravity is
\begin{eqnarray}
  \left.\frac{d\omega}{dt}\right|_\mathrm{self,jk}
&  = & \int_{\mathrm{ringlet}\; j} df
\left.\frac{d\omega}{dt}\right|_\mathrm{point} \nonumber \\
&  \simeq & -\frac{1}{\pi}\frac{h_km}{NM}\frac{\Omega}{e}\frac{a}{a_k-a_j}\,\frac{q_{jk}\left(1-\sqrt{1-q_{jk}^2}\right)}{q_{jk}^2\sqrt{1-q_{jk}^2}}
  \;\;\; .
\label{eq:grav1ring}
\end{eqnarray}
Since the differential precession rate of ringlet $j$ due to Chariklo's $J_2$ is
\begin{equation}
  \left.\frac{d\omega}{dt}\right|_{J_2,j}
  \simeq {\rm const} -\frac{21}{4}J_2\frac{R^2}{a^2}\frac{a_j-a}{a}\Omega
  \;\;\; ,
\end{equation}
the condition that self-gravity be strong enough to maintain apse
alignment against $J_2$-induced differential precession is
\begin{align}
\qquad\qquad
  P& = \left.\frac{d\omega}{dt}\right|_{J_2,j} + \sum_{k\neq j}
  \left.\frac{d\omega}{dt}\right|_\mathrm{self,jk} \, , \qquad 0<j,k\leq N\\
  &= C_1j + C_2m\sum_{k\neq j}\frac{h_k}{k-j}
  \label{eq:selfgrav1}
\end{align}
where $P$, the overall ring precession rate, is independent of $j$,
and $C_1$, $C_2$ are functions only of $M$, $J_2$, $a$, $\Delta a$,
$e$, $\Delta e$. In Eq.~\ref{eq:selfgrav1} we also take the
ringlets to be evenly spaced in semimajor axis, $a_k-a_j=(k-j)\Delta
a/N$, without loss of generality. With $e_1=e$, $e_N=e+\Delta e$, and
a choice for the ring density profile $h_k$, Eq.~\ref{eq:selfgrav1}
gives $N$ equations in the $N$ unknowns $P$, $m$, $e_2$, ...,
$e_{N-1}$ that in general have an exact solution.

\begin{table}
\begin{center}
\rule[.02in]{\textwidth}{.004in}
\begin{tabular}{llccc}
 Chariklo&  Equatorial radius$^1$ \makebox[.2in]{ } & $R$& \makebox[.2in]{ } & $144.9\pm 0.2$\\
&  Oblateness$^1$& $\varepsilon$& & $0.213$\\
& & \\
Inner ring \makebox[.2in]{ } & Radius$^1$& $a_1$& & $390.6\pm 3.3$\\
& Radial width$^2$& $\Delta a_1$& & 5.5 to 7.1\\
& Optical depth$^1$& $\tau_1$& & $0.449\pm 0.009$\\
& & & & $0.317\pm 0.008$\\
& & & & \\
Outer ring& Radius$^1$& $a_2$& & $404.8\pm 3.3$\\
& Radial width$^1$& $\Delta a_2$& & $3.6^{+1.3}_{-2.0}$\\
& Optical depth$^1$& $\tau_2$& & $0.05^{+0.06}_{-0.01}$\\
& & & & $0.07^{+0.05}_{-0.03}$\\
& & & & \\
\makebox[.5in][l]{Gap between the rings$^1$}& & & & $9.0\pm 0.4$\\
& & & & $8.3\pm 0.2$\vspace{-.12in}
\end{tabular}
\end{center}
\rule[-.02in]{\textwidth}{.004in}
\begin{small}$^1$\citet{bragaribas14}\hfill\\$^2$\citet{sicardy14,elmoutamid14}
\end{small}
\caption{Observed physical properties of 10199 Chariklo and its
  rings. All lengths are in kilometers. Where two values with
  uncertainties are listed, the first was measured from the ingress
  and the second from the egress portion of the 2013 June 3
  occultation.}
\label{table:chariklo}
\end{table}

\subsection{Results}\label{sec:results}

\begin{figure}
\begin{center}
\includegraphics[scale=0.35]{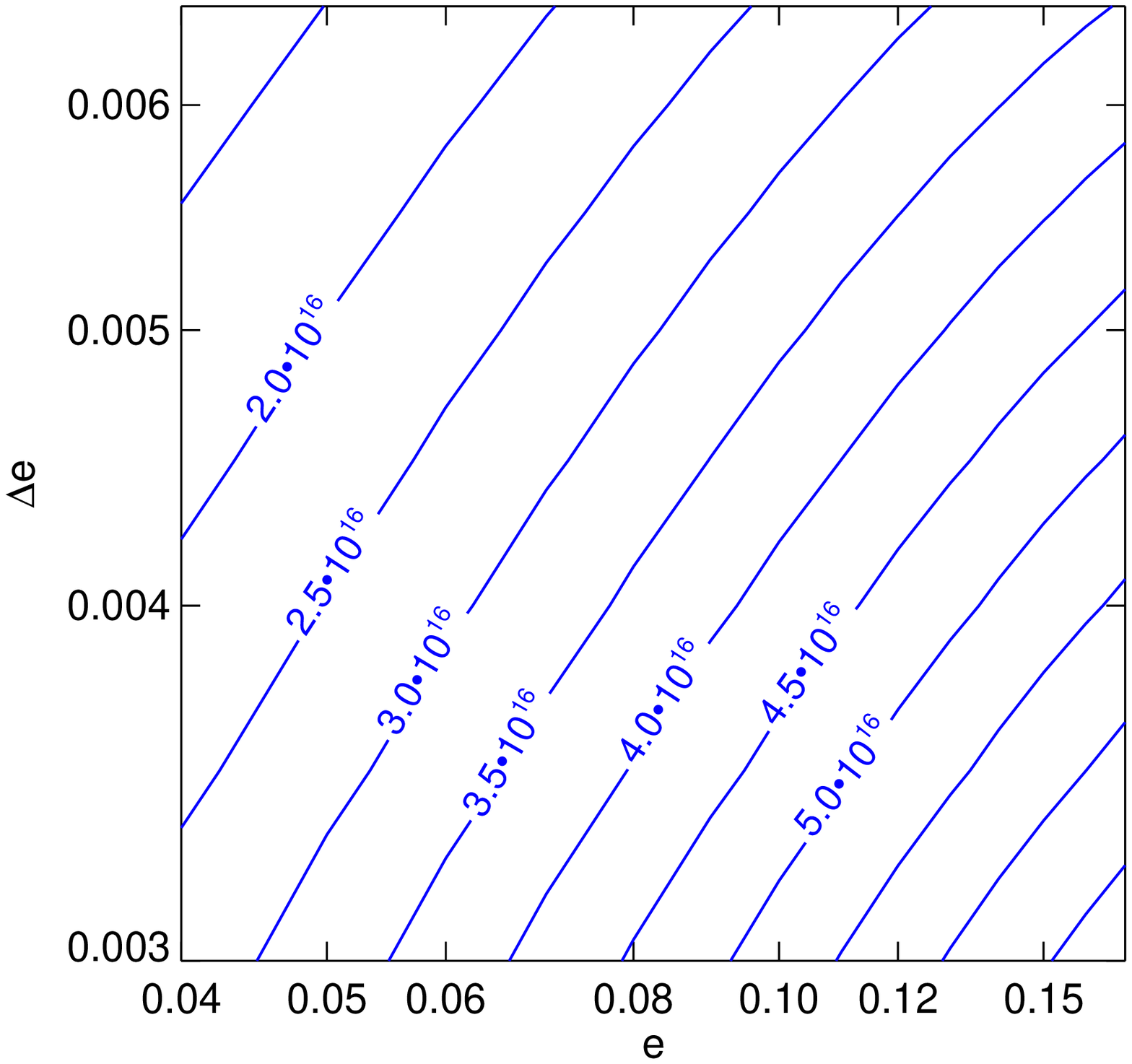}\\
\includegraphics[scale=0.35]{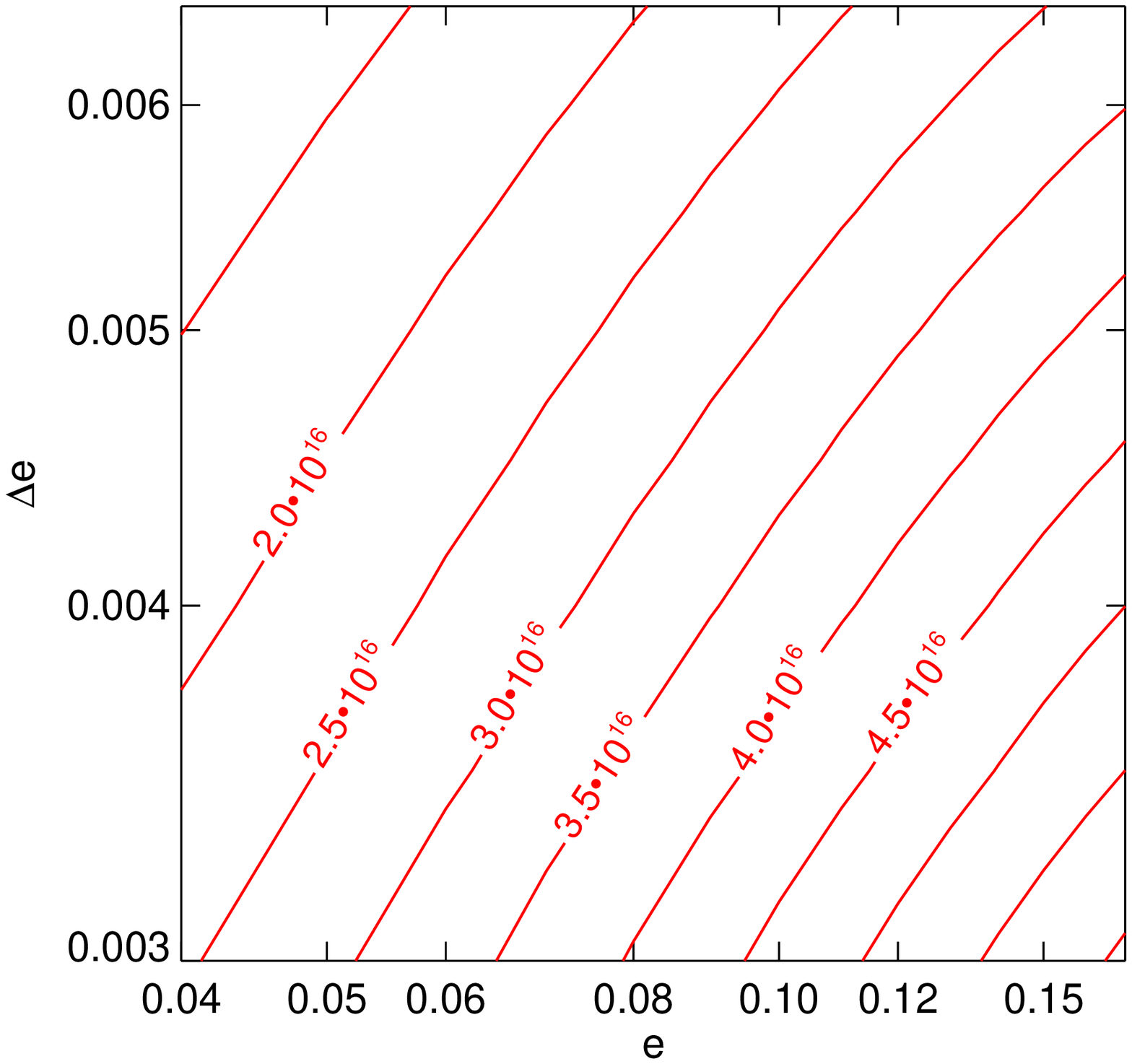}\\
\includegraphics[scale=0.35]{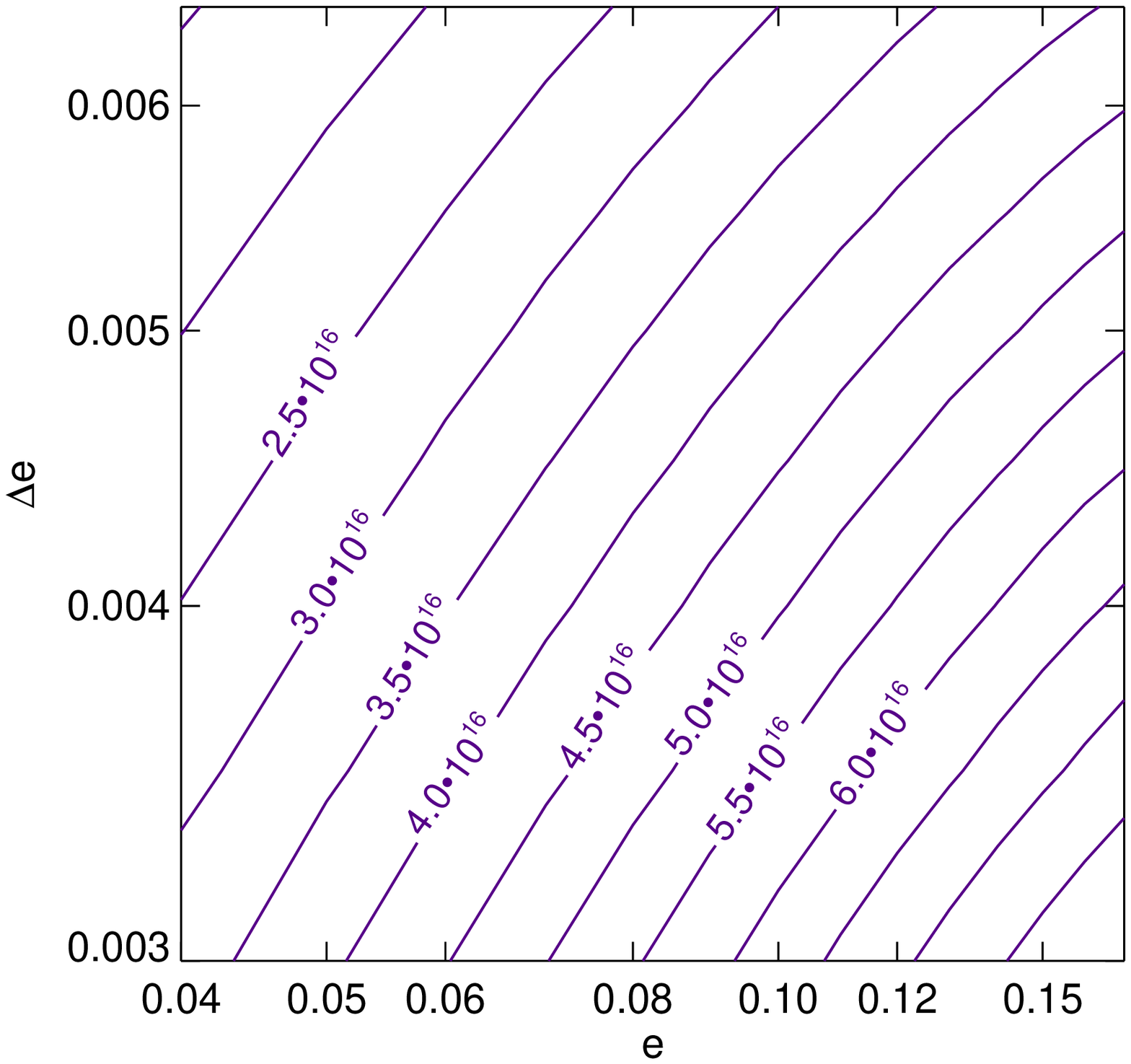} 
\end{center}
\caption{Contour plots of total ring mass in grams as a function of
  $e$, $\Delta e$ for three possible ring surface density profiles:
  uniform, $h_k=1/N$ (top); parabolic with edge densities half the
  value of the central density, $h_k=3(j/N)^2-3j/N+3/2$ (middle);
  parabolic with edge densities twice the value of the central
  density, $h_k=-12(j/N)^2/5+12(j/N)/5+9/5$ (bottom). The total mass
  increases by an order unity factor for profiles with mass
  concentrated towards ring edges. }
\label{fig:masscontour}
\end{figure}

Because we have only a rough lower bound on $\Delta e$ and $e$, we
solved the system of Eq.~\ref{eq:selfgrav1} for a grid of values
$0.06\leq e\leq 0.17$, $0.003\leq \Delta e\leq 0.0065$. By analogy to
the Uranian system, we chose $e\gtrsim \mathrm{several}$~times~$\Delta
e$. An $N$-dimensional Newton's method solver produced the ring masses
in Figure~\ref{fig:masscontour} for constant and quadratic surface
density profiles $h_k$. Further experiments with profiles of the form
$h_k = Ak^p + B$ with $p$ even, $0\leq p\leq 16$ and $A$ and $B$
coefficients chosen to yield $0.1\leq \max(h_k)/\min(h_k)\leq 10$,
gave ring masses differing from those of Figure~\ref{fig:masscontour}
by less than an order of magnitude. We performed similar experiments
varying Chariklo's shape and bulk density. Since the oblate spheroid
shape found by \cite{bragaribas14} is somewhat uncertain, and since
the rotation lightcurves of \cite{fornasier14} suggest Chariklo may
have an equatorial axis ratio of at least 1.1, we tried different
$J_2$ values representing a Jacobi ellipsoid Chariklo with axis ratio
1.1 to 1.6 in the ring plane.  Also, since densities for binary KBOs
of sizes similar to Chariklo are typically 0.6 to 0.7~g~cm$^{-3}$
\citep[see, for example,][]{grundy15}, we tried lowering our mass
estimate for Chariklo to reflect these densities.  These experiments
together gave ring masses differing from those of
Figure~\ref{fig:masscontour} by about a factor of two. This is
reasonable: as equations~\ref{eq:grav1ring}-\ref{eq:selfgrav1}
indicate, the ring mass is roughly proportional to Chariklo's mass and
$J_2$ moment, and these Jacobi ellipsoids have $J_2$ moments about
twice that of the nominal oblate spheroid.

In short, our model predicts
\begin{align}
  \mathrm{ring\;mass}\;m& \simeq \mathrm{few}\times 10^{16}\;\g\\
  \mathrm{average\;surface\;density}\;\sigma& \simeq \mathrm{few}\times 100\;\g\;\cm^{-2} \;\;\; .
\end{align}
Assuming the ring particles collide frequently and form a monolayer --- not unreasonable given the optical depth $\tau_1\sim 0.32-0.45$ --- we may also estimate
\begin{align}
  \mathrm{typical\;particle\;size}\;s& \simeq \mathrm{few}\;\m\\
\mathrm{typical\;random\;velocity}\;v& \simeq \mm/\s
  \;\;\; .
\end{align}
The ring appears slightly gravitationally unstable with
Toomre $Q\simeq$~a few tenths, and the particle size is
  coincidentally similar to that in Saturn's C ring \citep{cuzzi09}.  

We can also estimate a timescale for spreading due to collisional
diffusion by noting the time between collisions is the orbit period
divided by the optical depth, a typical collision changes a particle's
energy by a fraction $v/(\sqrt{3}\Omega a)$, and the potential energy
spread across the ring is $\Delta a/a$. This gives
\begin{equation}
\mathrm{radial\;spreading\;time} \sim \left(\frac{\Delta a/a}{v/(\sqrt{3}\Omega a)}\right)^2\frac{1}{\tau_1}\frac{2\pi}{\Omega} \sim 10^5 \;\mathrm{years} \;\;\;.
\end{equation}
In the above we have neglected precession due to collisions within the
ring since we believe that particle-particle collisions contribute
much less than self-gravity to differential precession for the ring
profiles we consider, in contrast to the situation in Uranus's
$\epsilon$ ring \citep{goldreich79,chiang00}. While both $J_2$ and
$\Delta a/a$ are much larger in the Chariklo system than in the Uranus
system, the $R/a$ and random velocity values are comparable. As a
result, precession from the central body's $J_2$ is much larger in the
Chariklo system than in the Uranian case, while collisionally induced
precession is much less important. However, if as in the work of
\citet{chiang00} on the Uranian system (a) shepherd satellite(s)
increase(s) the random velocity by more than an order of magnitude
near the edge(s) of Chariklo's ring, collision-induced precession at
the edge(s) may significantly affect the derived ring mass: to balance
the collisional and $J_2$ terms in the Uranian rings, the total ring
mass needs to increase by of order a factor of 10
\citep{chiang00}.

Massive shepherd satellites could also significantly lengthen the
spreading timescale. Confirming the existence of such (a) satellite(s)
would require either an extremely high resolution ring profile or
excellent luck in occultation timing. However, circumstantial evidence
suggests that the satellite(s), if present, would be no larger than
$\sim$1~km in size. To be consistent with the width of the gap between
Chariklo's inner and outer rings, a shepherd in the gap would need to
be $\sim$1~km in size. Any shepherd interior to the rings must also be
small. Since the rings are slithgly gravitationally unstable, with
Toomre $Q<1$, they must lie within Chariklo's Roche radius in order to
avoid fragmentation.\footnote{Both shepherds and a Jacobi ellipsoid
  shape for Chariklo would decrease the Toomre $Q$ further.} An inner
shepherd would likewise lie within the Roche radius. Since KBOs larger
than $\sim$300~km are expected to be rubble piles \citep[see, for
  example,][and references therein]{stewart09}, a large $\gtrsim 1$~km
inner shepherd would likely be subject to tidal disruption.  As the
total ring mass is comparable to that of a 1-km body, a 1-km shepherd
would slow collisional diffusion only by a factor of order
unity\footnote{A large $>$1~km outer shepherd paired with a small
  inner shepherd would not slow diffusion significantly, as the ring
  would preferentially spread inward}, at most mildly lengthening our
estimate for the spreading time.

\section{Constraints on dynamical formation scenarios}\label{sec:formation}

Recent reinterpretation of occultation data of the second-largest
Centaur, 2060 Chiron, as possible evidence of rings \citep{ortiz15,ruprecht15}
has further sharpened interest in ways to form rings around small
bodies which have had close encounters with giant planets. Dynamical
formation mechanisms proposed thus far for Centaur ring systems
\citep[see, for example,][]{bragaribas14,elmoutamid14,duffard14} fall
into two broad categories which we discuss in light of our findings in
\S\ref{sec:selfgravity}.

\subsection{Ejection}\label{sec:ejection}

First, the rings may have formed while the Centaur was still a member
of the Kuiper belt, probably via ejection of a small amount of
material off the parent body's surface into an orbit within the Roche
radius. Cratering collisions are one way to do this. Simulations of
such collisions at the $\sim$1~km/s velocities typical of the Kuiper
belt \citep[see, for example,][]{leinhardt12} indicate that the
ejected mass is of order the impactor mass and that the vast majority
of ejected material departs at speeds of order the escape
velocity. Simulations of collisional satellite formation \citep[see,
  for example,][]{canup04} likewise indicate that for velocities above
the escape velocity, $<$1\% of ejected material remains in orbit
long-term. So most likely only a small fraction of the ejecta in any
given collision with Chariklo (escape velocity $\sim$100~m~s$^{-1}$)
would remain in a close orbit. As the total mass of Chariklo's rings
is of order that of a $\sim$1~km body, a ring creation event would
require a substantially larger impactor which we take here to be
$\sim$5~km.

From \citet{schlichting12}, the optical depth to collisions for a
$\sim$150-km target is $\sim$$4.6\times 10^{-8}
[{5\km}/{\rm{projectile\;size}}]^{2.7}$, suggesting that a body like
Chariklo would have collided with a $\sim$5-km projectile a few times
over its few-Gyr residence time in the Kuiper belt. It therefore seems
plausible that most KBOs of Chariklo's size could acquire ring systems
or at least small satellites in this way.  However, the short
$\sim$$10^5$-year ring spreading time we find in \S\ref{sec:results}
suggests that such systems formed of order $\gtrsim$1 Gyr ago would
have dispersed to low optical depths by the present day.

Rotational disruption may also lift boulders from the surface into
orbit; it may explain how near-Earth asteroids \citep[see, for
example,][and references therein]{marchis08} and in at least one case
a KBO \citep{ortiz12} shed material from their surfaces into orbit,
creating small satellites. This
mechanism seems less likely for Chariklo because its $\sim$7-hr spin
period \citep{Fornasier} is not extremely close to breakup\footnote{Some binary asteroids that are believed to have formed
    via rotational fission do rotate slowly \citep[see, for
    example][]{pravec10}. This occurs when the secondary is massive
    enough to convert rotational energy of the primary into its
    orbital energy.}, and because rotational disruption is infrequent enough that occurrence in the last $10^5$ yrs is unlikely.


\subsection{Satellite disruption}\label{sec:satellite}

Second, the rings may have formed during the close encounter(s) with a
giant planet, most likely Neptune, that brought the parent body from
the Kuiper belt to a Centaur orbit. For example, this encounter may
have perturbed a small moon inward the Roche radius, forcing it to tidally
disrupt into a ring.

Such a moon may have been captured by dynamical friction in the early
history of the Kuiper belt \citep{noll08,goldreich02}. Moons of this
kind tend to have close orbits and be similar in mass to the central
body. Alternatively, as discussed above, a typical $\sim$150-km KBO
today has collided with a $\sim$5~km projectile, placing perhaps
$\lesssim$1\% of the projectile mass into orbit. If these ejecta end
up outside the Roche radius, they may well coalesce into
satellites as described by \citet{canup04} for the Earth-Moon system
and hypothesized for the small moons observed around large KBOs
\citep{brown06}.

The tidal force needed to perturb a satellite inside the KBO's Roche
radius during a Neptune close encounter depends on the satellite's
initial orbit. We consider the case where the semimajor axes of the
initial satellite orbit and the final ring differ by a factor of order
unity and justify below why this is the most probable scenario. If the
satellite's initial semimajor axis is $a_0$, the change in velocity produced by the tidal force during a single Neptune encounter
with closest approach distance $b$ is
\begin{equation}
\frac{GM_\mathrm{Neptune}}{b^2}\frac{2a_0}{b}\frac{b}{u}\, ,
\end{equation}
where $u$ is the typical random velocity of a Neptune-crossing KBO.
To change the semimajor axis by a factor of order unity, this velocity change must be of order the ring orbit velocity $\sqrt{GM/a_1}=\Omega a_1$, implying
\begin{equation}
  b\sim\sqrt{\frac{2 GM_\mathrm{Neptune}}{u\Omega}}
  \sqrt{\frac{a_0}{a_1}} \sim
  3.6\times 10^{10}\;\cm \sqrt{\frac{a_0/a_1}{2}}\sim 15\;R_\mathrm{Neptune}\, ,
\label{eq:beqn}
\end{equation}
where we use $u\simeq 2$~km/s. 
For this encounter, the impact parameter at
infinity is $b_\infty\sim b\sqrt{GM_\mathrm{Neptune}/b}/u\sim 7.8\times
10^{10}$~cm. The same encounter changes the KBO's solar orbital
velocity by of order $GM_\mathrm{Neptune}/(b_\infty u)\sim 4.4$~km/s,
which is just enough to move the KBO onto a Centaur orbit. The Centaur
creation event may thus also serve as a ring creation event.

Note that this scenario works only for KBOs that attain Centaur orbits
via a single close Neptune encounter rather than multiple weak
encounters. The much longer encounter duration in a weak scattering
means the satellite can complete many orbits during a single
encounter, severely diluting the effect of the tidal force. As a
result, the KBO will likely become a Centaur before the moon's orbit
changes significantly. At the same time, a single very close Neptune
encounter that changes the moon's orbit velocity by more than an order
unity factor will most likely unbind the moon from the KBO. Ring formation via satellite disruption thus requires a single Neptune encounter and a close match between the satellite orbit and the Neptune encounter strength.

This condition allows us to estimate the frequency of such satellite
disruptions.  Since the number of Centaurs created scales
logarithmically with the distance of the closest encounter, and there
are about $6$ natural logarithmic intervals between $10 R_{\rm
  Neptune}$ and Neptune's Hill radius, we expect that about one in six
Centaurs formed via a single Neptune encounter. The typical Centaur
dynamical lifetime of $\sim$few million years \citep{bailey09} and the
$\sim$$10^5$-year ring spreading time suggests that only a small
fraction of these, $\sim 10\%$, would remain visible today. In the
best scenario --- i.e.~if every large KBO has a close-in satellite,
and assuming the single encounter with Neptune is just strong enough
for the KBO to attain a Centaur-class orbit --- the chance of
observing a Centaur with rings is $\sim$few percent.

A further difficulty is the rapid rate of collisional destruction of
small moons.  At an impact velocity of $\sim$1~km/s, the typical
random velocity in the Kuiper belt, a $\sim$20~m impactor can destroy
a $\sim$1~km moon. Assuming a size distribution consistent with
\cite{schlichting12}, the rate at which a given 1~km target encounters
20~m impactors in the current Kuiper belt is about once every $10^5$
years, far faster than the creation rate of 1~km moons by impacts as
described in \S\ref{sec:ejection} above. Rotational fission is
likewise unlikely to create moons fast enough to beat collisional
destruction.

The fast collisional destruction rate also makes significant
collisional perturbation of a moon's orbit difficult. While a
projectile of size at least $\sim$few hundred meters is needed to
change the momentum of a 1~km moon by a factor of order unity, such a
collision is likely to occur only once every $\sim$Gyr. Even assuming
the moon's orbit is just a factor of $\sim$2 larger than the Roche
radius, the moon will be destroyed before it can be collisionally
moved close enough to the Centaur.

For satellite disruption to be viable for ring formation at all, the
satellites must therefore be more massive than the observed
ring. Since the catastrophic collision timescale for a 10- to 20-km
KBO is about a Gyr, a moon of that size could plausibly survive until
Centaur orbit insertion. The fraction of $\sim$100-km KBOs with moons
of this size is unknown: these moons are too large to form efficiently
via cratering impacts and too small to be captured efficiently via
dynamical friction, though rotational fission may provide an effective
pathway. Nevertheless, moons of similar size have been found around a
few of the largest KBOs \citep{ragozzine09,brozovic15}. Tidal
disruption of such a large moon inside the Roche radius would lead
initially to a much more massive disk/ring, but because the initial
diffusive spreading time would be correspondingly shorter, the ring
lifetime would not be significantly longer than our estimate in
\S\ref{sec:results} above.

We conclude that among dynamical channels, compatibility between the
ring spreading timescale and the expected time since formation weakly
favors formation during a Neptune close encounter, not during the
few-Gyr residence time in the Kuiper belt. However, this scenario
requires that most $\sim$100-km KBOs have moons of size tens of
kilometers or more, and it predicts a Centaur ring occurrence rate of
at most a few percent.

\section{Ring formation via outgassing}
\label{sec:outgas}

In view of the uncertainties in the dynamical formation channels
discussed in \S\ref{sec:formation}, we develop a contrasting
mechanism for ring formation, dusty outgassing, briefly alluded to by
\citet{bragaribas14}. Specifically, we discuss a simple model in which
a dusty outflow entrained by CO sublimating within Chariklo deposits
particles in close orbits. In \S \ref{subsec:confront}, we collect
  the available Centaur observations and discuss them in the context
  of our theoretical predictions.

\subsection{Chariklo's history of CO loss}
\label{subsec:co}

We first estimate the rate and velocity of CO outgassing from
Chariklo, taking into account its dynamical history. We focus on CO
here because of its high abundance and physical-chemical properties,
but most of the discussion below applies if another volatile gas
e.g. N$_2$, CO$_2$, CN, is substituted. In particular, N$_2$ has
properties very similar to CO in both triple point temperature
  and saturation vapor pressure and may be more abundant.

We do not consider the alternative outgassing model where volatile
gases are released by the exothermic process of water ice
crystallization. While this process would increase the mass-loss rate
above our estimates, we believe the correction would be order unity or
less: scaling the crystallization model for Chiron \citep{Prialnik} to
the distance of Chariklo gives an outgassing rate comparable to or
smaller than ours.

\subsubsection{CO free escape vs. diffusion}\label{sec:tworates}

As previous works have noted \citep[see, for
  example,][]{Cowan,Jewitt}, sublimation of surface CO is
fast even for bodies in the Kuiper belt. An estimate of the maximum sublimation rate follows from equating the rates of insolation and latent heat absorption:
\begin{eqnarray} 
{\dot m}_{\rm CO,max} & \leq & \frac{1}{E_{\rm latent}}
  {\frac{(1-A) L_\odot}{4 \pi a^2}} {\frac{\pi R^2}{4\pi R^2}}
  \nonumber \\ & \sim &
  5.3\times 10^{-7}
  \g/\s/\cm^2 \, \left( \frac{a}{16 \AU}\right)^{-2}\, ,
\label{eq:mdot}
\end{eqnarray}
where the latent heat of CO sublimation is $E_{\rm latent} \sim
240\J/\g$, the albedo is taken to be $A \sim 0.04$ \citep{Fornasier},
and $a$ is Chariklo's semimajor axis in its orbit around the
sun. Including surface cooling introduces only a
small correction \citep{Cowan,Jewitt}.  At this rate, assuming a CO
mass fraction $f_\mathrm{CO}\simeq0.10$ \citep[in line with
  previous comet measurements in e.g.][]{previous}, Chariklo would have
outgassed all of its CO in $10^5$ yrs in the Kuiper belt \citep[also
see][]{Jewitt}, well before being scattered inward.

However, as numerous works have also argued
\citep[e.g.,][]{Enzian,Sanctis,guilbert}, this rate applies only to CO
that is exposed on the surface and can escape freely.  If CO is
instead interspersed with ice and dust grains throughout the
body, then once the surface CO departs, the CO vapor must first
diffuse upward through the ice/dust matrix in order to
escape\footnote{In principle, wherever the vapor pressure exceeds the
  tensile strength of the porous grains, vents can open and reexpose
  CO. However, even with the tensile strength $\sim 10^4
  \dyne/\cm^{-2}$ measured from tidal splitting of comets and lab
  simulations of cometary ice \citep{Donn,Whipple,Kochan}, the CO
  vapor pressure (see Fig.~\ref{fig:CO}) is too low to form vents at
  any significant depth.} and the loss rate slows significantly.
 
To estimate this slower loss rate, we consider the upward diffusion of
a CO molecule through convoluted tunnels formed by loosely packed
icy/dusty grains.\footnote{The vapor density is so low that gas-gas collisions
  are irrelevant and the so-called 'Knudsen regime' applies.}  Let the
average size and number density of grains be $s$ and $n$ respectively.
The porosity of the medium is then $\phi = 1-n 4\pi/3 s^3$ and the mean free
path for gas-grain collisions is $l_{\rm mfp} \sim 1/n \pi s^2 \sim
s/(1-\phi)$.  Assume solid CO remains only below a depth $\Delta
\ell$. The CO vapor pressure is the saturated value $p_v = p_v (T)$ at
this depth and decreases upward.  The diffusive mass flux driven by
the vapor density gradient over a single tunnel is
\begin{equation}
J \sim \pi s^2 D \frac{d}{dr} \rho_v \sim \pi s^2 D
\frac{p_v(T)}{c_s^2 \Delta \ell}\, ,
\label{eq:flux}
\end{equation}
where we have assumed the typical tunnel cross section is $\pi s^2$,
the entire body has temperature $T$, $c_s$ is the ideal gas sound
speed at $p_v$ and $T$, the vapor mass density is $\rho_v$, and the diffusion
coefficient is \citep{Evans}
\begin{equation}
D \sim \frac{\phi}{\tau}  l_{\rm mfp}\, c_s \sim \frac{\phi}{\tau} \frac{s}{1-\phi} c_s\, ,
\label{eq:Ddiff}
\end{equation} 
where $\tau$, the so-called tortuosity, covers our ignorance on
factors such as the extended length of a tunnel due to topology, the
width distribution among tunnels, and the scattering property of gas
molecules off the tunnel wall.
Over a unit surface area containing $\sim$$1/(\pi s^2)$ tunnels, the
total mass-loss rate is
\begin{equation}
{\dot m}_{\rm CO}  \sim \frac{1}{\pi s^2} J \sim \frac{\phi}{\tau}
\frac{s}{1-\phi} {\frac{p_v(T)}{c_s \Delta \ell}}\, .
\label{eq:diffusion}
\end{equation}

To determine the appropriate depletion depth $\Delta \ell$, also
called the sublimation front, we assume that Chariklo has been
outgassing at temperature $T$ for most of its lifetime $t_\mathrm{age}$,
\begin{equation}
{\dot m}_{\rm CO} \times t_\mathrm{age} = \Delta \ell 
\times \rho \times f_{\rm CO}\, .
\label{eq:ell}
\end{equation}
We take the saturation vapor pressure to be \citep{Clayton}
\begin{equation}
p_v(T) \simeq  0.15 {\rm \, bar}\, \exp\left[ 6.15\left(1-
    {\frac{68\K}{T}}\right)\right]\, , 
\label{eq:pvapor}
\end{equation}
where the temperature is the local blackbody value at distance $a$,
that is, $T=T_\mathrm{BB}$ where
\begin{equation}
\sigma_\mathrm{SB} T_\mathrm{BB}^4 = {\frac{(1-A) L_\odot}{16 \pi a^2}}\, .
\label{eq:mdot2}
\end{equation}
This assumes that the heat needed for sublimation is negligible
compared to re-radiation (as we easily verified).  The CO loss rates
from this simple model, shown as red dots in Fig.~\ref{fig:CO}, are
orders of magnitude below the free escape loss rates (solid lines). In
particular, in the $\sim$few~Gyr that Chariklo presumably spent at
40~AU, the CO sublimation front would reach a depth 1~km \citep[also
see][]{Sanctis}.\footnote{The limited loss of CO during the KBO
    stage also explains why objects like comets 29P/SW1 and comet
    Hale-Bopp showed CO outgassing. See \S \ref{subsubsec:minting}. } If
Chariklo had instead spent that time at 20~AU, all its CO would have
been depleted.

Interestingly, we reach nearly identical conclusions if we
  substitute N$_2$ for CO in the above discussion. This suggests that
  evaporation of CO or N$_2$, or both, can drive dusty outgassing. In
  contrast, CH$_4$ and CO$_2$ areunimportant here due to their much
  higher triple point temperatures.

\subsubsection{Becoming a Centaur}
\label{subsubsec:minting}

Chariklo's arrival in its current orbit ($a=15.8$~AU)
$\sim$few Myrs ago \citep{bailey09} should have strongly increased its
temperature and CO loss rate, as we discuss below \citep[also see][]{Sanctis2}.

\begin{figure}
\begin{center}
\includegraphics[scale=1.4]{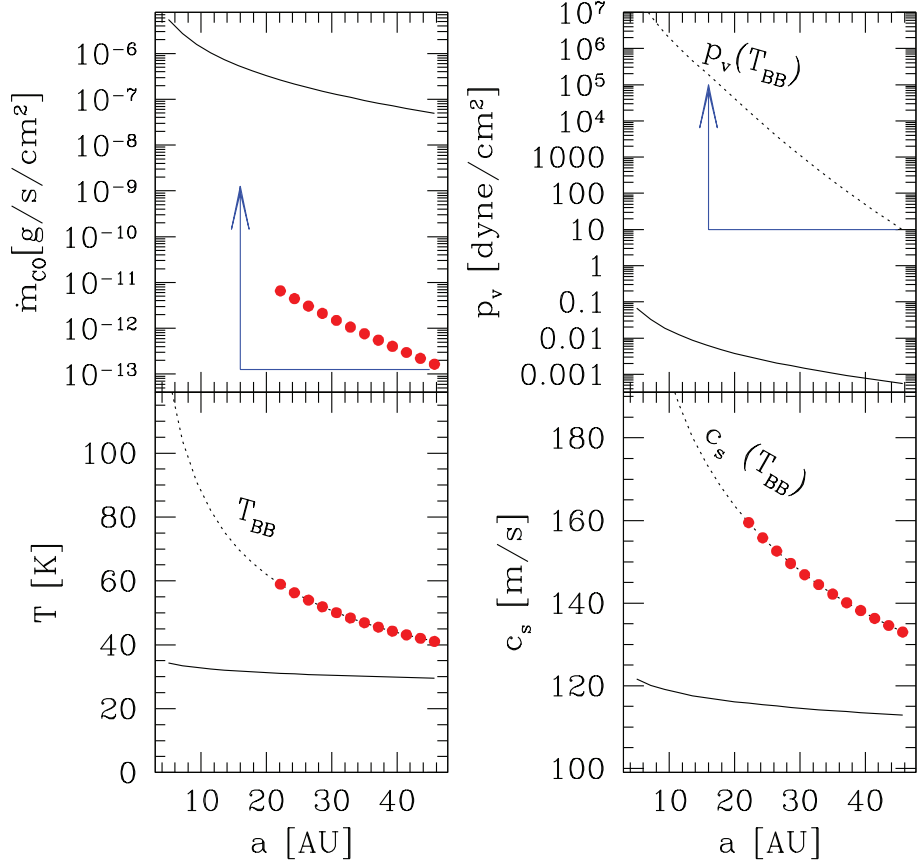}
\end{center}
\caption{Comparison of (top left) CO outgassing rate, (bottom left)
  temperature, (top right) CO saturation vapor pressure, and (bottom
  right) sound speed under three different models. The black solid
  curves show conditions when CO is exposed on and sublimating
  directly from Chariklo's surface (Eq.~\ref{eq:mdot}).  In this
  model, most of the solar insolation is channelled into latent heat
  and the body maintains a temperature lower than the local blackbody
  value (black dotted curves marked $T_\mathrm{BB}$). The red dots
  show conditions when surface CO is absent and CO loss occurs only
  through upward diffusion through an ice/dust matrix
  (Eq.~\ref{eq:diffusion}). In this model, Chariklo stays close to the
  local blackbody temperature. For the computations we use bulk
  density $\rho= 1$~g~cm$^{-3}$, grain size $s = 1\,\mu$m, porosity
  $\phi=0.5$, tortuosity $\tau=2$, $t_\mathrm{age} = 4.5$~Gyr and
  present only results where the depletion length $\Delta \ell$ has
  not yet reached Chariklo's center.  The blue arrows indicate the
  increases in pressure and CO mass loss when Chariklo abruptly
    moved from 40~AU to 16~AU on becoming a Centaur. Seasonal
  freeze-thaw cycles can cause local outbursts whose outgassing rates
  greatly exceed the mass-loss rate at the tip of the blue
    arrow.}
\label{fig:CO}
\end{figure}

While the surface temperature would have jumped immediately from
$\sim$41~K to 68~K once Chariklo reached 15.8~AU, the CO vapor
pressure would have remained low until the CO-rich matrix at depth
$\Delta\ell\simeq 1$~km also warmed. To estimate the heat diffusion
time through $\Delta\ell$, we take the thermal conductivity to be
$\kappa \sim 0.1$~W~m$^{-1}$~K$^{-1}$, appropriate for a dust-ice
aggregate \citep{Enzian,Sanctis,guilbert}, and we use specific heat
capacity $c_p \sim 6\times 10^6$~erg~g$^{-1}$~K$^{-1}$ for an
ice-dominated mixture between $40$ and $70$ Kelvin
\citep{Klinger}. The thermal diffusivity is then $\nu = \kappa/\rho
c_p \sim 10^{-3}\,\cm^2\,\s^{-1}$, so the CO-rich interior should have
stayed cold for $\sim$1~Myr after Centaur orbit insertion.  After this
delay, the CO vapor pressure in the interior would have increased by
a factor of $\sim 10^4$ (eq.~\ref{eq:pvapor}). The CO diffusion rate
would likewise have increased (eq. \ref{eq:diffusion}) from the
original $10^{-13}\,\g\,\s^{-1}\cm^{-2}$ to $\sim
10^{-9}\,\g\,\s^{-1}\,\cm^{-2}$, or a global rate of $
10^7\,\g\,\s^{-1}$ (also see
  Fig. \ref{fig:COlimit}).\footnote{This is far above the result of
  \citet{Sanctis2} for a similar object. We suspect the difference
    arises because their integration is too short to allow the
  interior to thermalize.} These jumps are shown as blue arrows in
Fig.~\ref{fig:CO}. The energy needed to maintain the new rate is
negligible compared to insolation, consistent with our assumption in
\S\ref{sec:tworates}. Note that due to the sudden transition, this
rate is higher than the equilibrium rate, i.e. the rate we would
expect if Chariklo had remained at its current location for a few
Gyrs.

We now estimate the total CO mass sublimated. Thermal diffusion
limits the CO mass lost in Chariklo's time as a Centaur: since $D >
\nu$, new CO vapor can escape sufficiently fast and the thermal
diffusion and CO sublimation fronts should nearly coincide. After
10~Myr, the thermal diffusion front lies $\sim 5\km$ below the
surface, and the total CO lost is
\begin{equation}
\Delta m_{\rm CO} \sim \rho f_{\rm CO} \times 5\km \times
4\pi R^2 \sim 5\times 10^{19} \g \, .
\label{eq:mtot}
\end{equation}
This is a few orders of magnitude higher than the ring mass we
inferred, so the total CO outgassed is potentially
enough to lift a ring's worth of solids off the surface.

\subsubsection{The ``spring thaw''}

Since neither of the pole solutions published by \citet{bragaribas14}
coincide with Chariklo's orbit normal, we expect strong seasonal
variations in solar insolation. We now estimate the corresponding
  variations in CO outgassing for later reference in
  \S\ref{subsec:dustring}-\ref{subsec:confront}.  In Chariklo's polar
regions the insolation variations cause large temperature
fluctuations to a depth of order $d \sim \sqrt{P_\mathrm{orb}\nu} \sim
10 \m$, where $P_{\rm orb} = 62\;{\rm yrs}$ is Chariklo's orbit
period.\footnote{Chariklo's small eccentricity has little impact on
  temperature.}  The CO diffusing through this layer experiences
freeze-thaw cycles that drastically modulate the surface outgassing
rate.

Because CO spends $d^2/D \sim 2\times 10^6\s$ passing through this
layer, its volume density there is $\rho_\mathrm{CO} \sim {\dot
  m}_{\rm CO}\times (d^2/D) /d \sim 10^{-6} \,\g\,\cm^{-3}\, ({\dot
  m}_{\rm CO}/10^{-9})$, equivalent to a vapor pressure of $p \sim
2\times 10^2 \,\dyne\,\cm^{-2}$. This is the saturation vapor pressure at
$\sim 47\,\K$ (eq. \ref{eq:pvapor}), so if the winter hemisphere cools
below this point, CO will recondense. We can estimate the winter
temperature by assuming that all heat content stored in layer $d$ is
radiated as blackbody radiation during the winter\footnote{The latent
  heat released by CO condensing at a rate $10^{-9}\g/\s/\cm^2$ is
  negligible.}:
\begin{equation}
d\times c_p \rho_\mathrm{CO} (T_{\rm BB} - T_{\rm winter}) = \sigma T_{\rm winter}^4
\times \frac{P_{\rm orb}}{2}
\;\longrightarrow\;
T_{\rm winter} \sim 38 \K \, .
\end{equation}
Within the winter polar circle, CO therefore re-freezes on its way up
from the deeper warm layer instead of escaping; in one winter ${\dot
  m}_{\rm CO} \times {P_{\rm orb}/2} \sim 1\g/\cm^2$ accumulates,
mostly in the coldest layers near the surface.  When spring arrives,
the sublimation and escape of this trapped CO proceeds nearly as for
exposed CO (solid lines in Fig.~\ref{fig:CO}), limited only by the
  insolation energy budget (Eq.~\ref{eq:mdot}).  As this maximum
  rate, $\dot{m}_\mathrm{CO,max} \simeq 5\times 10^{-7} \g/\s/\cm^2$,
  is one or two orders of magnitude higher than the annual rate, the
  trapped surface CO is exhausted within a small fraction of an
  orbit. By assuming that the average annual CO loss is outgassed at
  the maximum rate, we estimate the minimum duration of such an
  outburst is $\sim$50 days. The true duration is likely longer.  Such
  short outgassing outbursts, occurring twice an orbit after equinoxes
  but not necessarily near periapse, may have been seen in the 2-D
  simulations of \citet{Enzian}.

\subsection{Dust lifting and ring formation}
\label{subsec:dustring}

Although the total CO lost is large compared to our ring mass of
$\sim$$10^{16}$~g, that does not guarantee that the momentum and
energy of the CO outflow are enough to entrain dust particles and lift
them into orbit.  The minimum flow rate needed to lift a grain of size
$s_\mathrm{dust}$ off the surface of Chariklo is
\begin{eqnarray}
& & 
 \frac{\rho_{\rm dust} \frac{4\pi}{3}s_\mathrm{dust}^3
      \frac{GM}{R^2}}{ \pi s_\mathrm{dust}^2 c_s}
  \sim 2\times 10^{-8}\, \g\,\cm^{-2}\,\s^{-1} \nonumber \\
& & \, \times \left(\frac{s_\mathrm{dust}}{1\,\mu \mathrm{m}}\right)
\left({\frac{170\,{\m\,\s^{-1}}}{c_s}}\right) \left(\frac{R}{120\,\km}\right)
\left({\frac{\rho}{0.5\, \g\,\cm^{-3}}}\right) \, ,
\label{eq:dustlift}
\end{eqnarray}
where we have adopted a dust grain density of $\rho_{\rm dust} = 1
\,\g\,\cm^{-3}$. While this assumes the gas-dust coupling is in the Stokes
regime, the large mean free path we expect in the tenuous CO flow
would reduce the drag by of order unity and increase the minimum
required CO outflow rate by a similar factor.

The average gas outflow rate of $10^{-9} \g\,\cm^{-2}\,\s^{-1}$
that we estimate in \S~\ref{subsubsec:minting} for a newly
  migrated Chariklo is thus unlikely to produce a dusty coma.  This
is consistent with Chariklo's observed inactivity
\citep{Fornasier}. However, our discussion of the seasonal CO
freeze/thaw suggests that in its current location, Chariklo could show
short bursts of activity with regional mass outflows exceeding the
constraint in Eq.~\refnew{eq:dustlift}.\footnote{The recent detection
  of diurnal cycle of water ice sublimaton on Comet 67P \citep{67P} is
  an exact analogy of this time-dependent process, but happens on the
  diurnal timescale.}  At these times, we expect the CO vapor to
approximately share its momentum with any entrained grains,
accelerating them to a final speed
\begin{equation}
v_{\rm dust} \sim \frac{c_s}{1 + f}\simeq \frac{170\,\m\,\s^{-1}}{1+f} \left(\frac{T}{68\,\K}\right)^{1/2}
\label{eq:soundspeed}
\end{equation}
where $f$ is the solid to gas mass ratio in the outflow.
If the dust and CO mass fractions are comparable, most dust grains
will not attain the escape velocity of $\sim$100~m/s --- they
remain bound to Chariklo.  Indeed, \citet{Meech} reached a similar
conclusion for Chiron, a slightly smaller Centaur, arguing that
predominantly bound grains may explain why Chiron's coma has a steep
surface-brightness profile more consistent with an atmosphere than an
unbound halo.

Initially the outflow puts grains on roughly radial trajectories that
 re-intersect Chariklo's surface within less than an orbit.
 Though Chariklo's rotation \citep[7-hr period;][]{Fornasier} imparts
 to the grains a tangential velocity of $\sim$one-third the surface
 escape speed (for a bulk density of $1\g\,\cm^{-3}$), this is only a
 fraction $\simeq 0.3 \left({{7\;{\rm hrs}}/{P_{\rm
       rot}}}\right)\left({{390\,\km}/
   a_1}\right)^{1/2}\left(\rho/(1\g\,\cm^{-3})\right)^{-1/4}$ of the
 orbital angular momentum at distance $a_1$.  Grains will therefore
 quickly re-accrete onto Chariklo unless they collide with other
 grains before re-intercepting the surface. We can estimate the
 collision probability as follows. The mean-free path of grains in the
 outflow is $l_{\rm mfp} \sim 1/n_{\rm dust} \pi\s_{\rm dust}^2$,
 where grain number density $n_{\rm dust} = f {\dot m}_{\rm CO}/v_{\rm
   dust}/(4\pi/3 s^3 \rho_{\rm dust})$. After a free-fall time
 ($1/\sqrt{G \rho}$), the collision probability is
\begin{eqnarray}
  \tau_{\rm collision}  & \sim & {\frac{t_{\rm dyn}}{l_{\rm mfp}/v_{\rm dust}}}\\
  & \sim & 5
  \left({\frac{f {\dot m}_{\rm CO}}{10^{-7}\, \g\,\cm^{-2}\,\s^{-1}}}\right)
  \nonumber
  \left({\frac{0.5 \,\g\,\cm^{-3}}{\rho}}\right)^{1/2} \left({\frac{1\,\mu
      \mathrm{m}}{s}}\right) \, ,
\end{eqnarray}
Each grain should suffer multiple collisions given our estimated flow
rate.\footnote{Also,
  once an optically thick ring forms, it should accrete grains
  efficiently.}

We expect collisions to tend to redistribute angular momentum among
grains, allowing a fraction $\epsilon_2$ of the grains to remain in
orbit. $\epsilon_2$ is hard to estimate as it depends on the detailed
collision geometry and collision probability.  However, these
surviving grains' final orbits are most likely close to the surface,
where collisions are more frequent and the required angular momentum
is lower.  Finally, grains that remain in orbit collide with one
another, damping their relative velocities. Since the grains inherit
Chariklo's rotational velocity when they leave the surface, their
total angular momentum is parallel to Chariklo's. The grains' mutual
collisions cause them to settle into Chariklo's equatorial plane in a
ring configuration.  This mechanism thus favors the formation of a
ring that lies inward of the Roche radius, though close-in moons
slightly further away are also possible. How these orbiting grains
might attain the narrow dense ring geometry currently observed around
Chariklo is not clear, however.

We can estimate the dust mass remaining in orbit as
\begin{eqnarray}
  m_{\rm ring} &\approx & f \epsilon_1 \epsilon_2
  \Delta m_{\rm CO} \sim 5\times 10^{16}\, \g \nonumber \\
  & & \times
  \left(\frac{f}{1}\right)
  \left({\frac{\epsilon_1}{0.1}}\right)
  \left({\frac{\epsilon_2}{10^{-2}}}\right)
  \left({\frac{\Delta m_{\rm CO}}{5\times 10^{19}\,
        \g}}\right) \, .
\label{eq:mring}
\end{eqnarray}
Here $\epsilon_1$ is the fraction of the CO loss that occurs in short
outbursts --- roughly, the fractional area of polar regions on
Chariklo multipled by the fractional length of winter. When the
obliquity is $90\deg$, $\epsilon_1 \sim 0.25$; the pole solutions of
\citet{bragaribas14} and the orbit given for Chariklo in the JPL Small
Body Database indicate the polar regions cover about one-half the
total surface, so we adopt $\epsilon_1\sim 0.1$. While $\epsilon_2$ is
highly uncertain, unless $\epsilon_2 \ll 10^{-2}$ the dust mass placed
in orbit via CO outgassing is comparable to our calculations in \S
\ref{sec:selfgravity}, which is encouraging.

The particle size of $\sim$few~meters we estimate for the current ring
is likely unrelated to the original grain size. Frequent collisions
may cause the grains to stick together mechanically, increasing the
typical particle size. We leave a detailed description of ring
  particle growth, and of the particles' settling into a narrow ring
  geometry, to future work.

\subsection{Confronting Observations}
\label{subsec:confront}

We now collect available observations on Centaurs to test two
important aspects of our model: 1) that CO should be preserved in the
deep interior and may sublimate and diffusively migrate to the surface
once the Centaurs are heated in their new location; and 2) that CO
outgassing may lift dust grains off their surfaces either continuously
for hotter Centaurs or upon the start of spring in colder Centaurs.

\begin{figure}
\begin{center}
\includegraphics[scale=0.8]{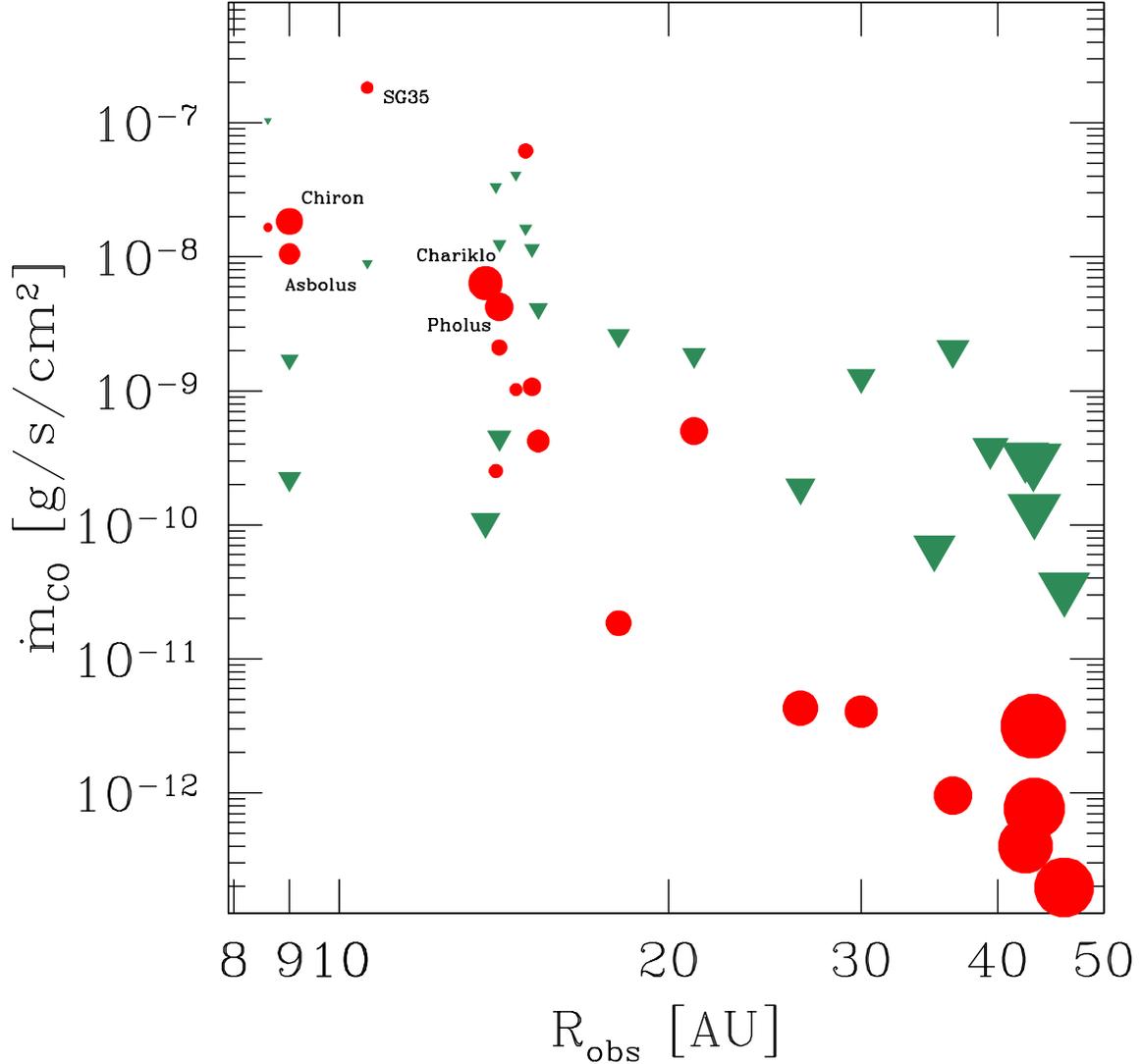}
\end{center}
\caption{Per-area CO outgas rates vs. heliocentric distance at
  measurement dates for various Centaurs and KBOs. The upper limits
  (green down-ward pointing triangles) are taken from
  \citep{RM,Jewitt08}, and the theoretical rates (filled circles) are
  calculated using eq. \ref{eq:diffusion}, assuming a temperature that
  is appropriate for a blackbody receiving the orbit-averaged
  illumination, an albedo of $A=0.05$, a porosity $\phi=0.5$, a
  tortuosity $\tau=2$, and a CO depletion depth of $\Delta \ell =
  1\km$. The theoretical predictions are inconsistent with the observed
  upper limits for a few large Centaurs, marked by their respective
  names. This could reflect a longer ($\gtrsim$few~Myr) residence time
  in the Centaur region. The symbol sizes indicate body sizes.}
\label{fig:COlimit}
\end{figure}

Highly relevant for the first of these are observations of comet
  outgassing, which suggest that CO (and possibly CO$_2$) is
  relatively abundant \citep[$\sim$10\% relative to
    H$_2$O,][]{previous}, while O$_2$ occurs at a surprisingly high
  $\sim$3\% \citep{bieler15,O2b} and N$_2$ occurs at a surprisingly
  low $\sim$0.1\% \citep{N2}. This motivates us to focus on CO.
We argue in \S\ref{sec:tworates} that while exposed CO on KBO surfaces
would have sublimated long ago, CO should retain its primordial
abundance in layers deeper than $\sim$1~km and could diffuse upward
rapidly upon displacement of the KBOs to warmer regions of the solar
system.  This may explain the detection of CO outgassing in Comet
29P/SW1 \citep[$a=6$~AU, $R\sim 15
  \km$,][]{Senay,Crovisier,Gunnarsson},
Comet Hale-Bopp \citep[at $a\sim7$~AU, $R \sim 20$~km,][]{Biver},
and possibly Centaur 2060 Chiron ($a=13.7$~AU, $R\sim 100\km$,
\citet{Womack}, though this has been challenged by
\citet{Rauer,BM}).  For these objects our simple model predicts CO
  loss rates between $10^{-7} \g\;\cm^{-2}\s^{-1}$ and $10^{-9}
  \g\;\cm^{-2}\s^{-1}$, depending on their current orbits, consistent
  with observations. But our explanation is not unique
    \citep[e.g.][]{Prialnik87}, and a number of surveys
    \citep{Rauer,BM,Jewitt08} have failed to detect CO around other
    Centaurs and KBOs.  Some of these upper limits contradict our
    predictions (Fig.~\ref{fig:COlimit}): for the few largest Centaurs
    like Chariklo, Chiron, Pholus and Asbolus, we predict outgassing
    rates some 1-2 orders of magnitude above the observed upper
    limits. This could be partly due to a residence time in the
    Centaur region of $\gtrsim$few~Myr, since CO outgassing slows as
    the sublimation front retreats further below the surface. 

\begin{figure}
\begin{center}
\includegraphics[scale=0.8]{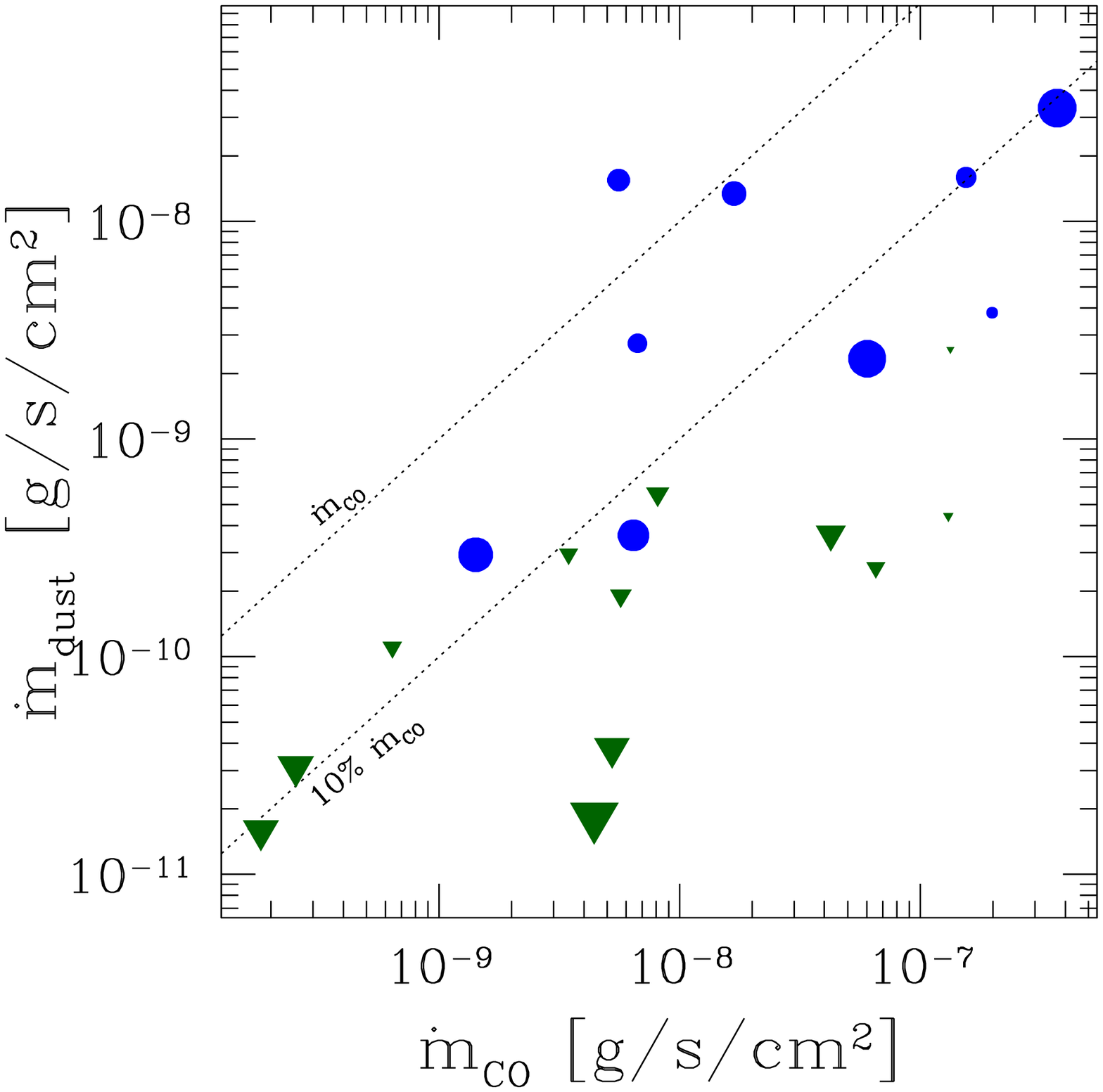}
\end{center}
\caption{Observed dust production rate vs.~predicted CO outgassing
    rate, with active Centaurs marked with filled circles and inactive
    ones marked as upper limits \citep{Jewitt}. To obtain the dust
    production rate per unit area, we use estimated radii from
    \citet{Jewitt}. Particularly for active Centaurs the radii may be
    overestimated; this would tend to produce underestimates of the
    dust production per unit area. Consistent with
    Eq.~\refnew{eq:dustlift}, nearly all active objects have CO outgas
    rates above $\sim$$10^{-8}\;\g\,\s^{-1}\cm^{-2}$. Moreover, the
    dust production rate roughly scales with the predicted CO
    rate. Likewise, most inactive objects have CO outgas rates below
    $\sim$$10^{-8}\;\g\,\s^{-1}\cm^{-2}$; however, a few Centaurs with
    large predicted CO outgas rates remain inactive.  }
\label{fig:activity}
\end{figure}

For information on dust grain lifting we turn to observations of
cometary activity in Centaurs \citep{Jewitt}. The origin of these cold
objects' activity remains a puzzle. \citet{Jewitt} has argued that CO
outgassing cannot be the correct explanation based on the simple
$1/r^2$ law expected for surface CO sublimation rate.  However, as our
calculation in Fig.~\ref{fig:activity} shows, our predicted CO outgas
rates largely explains the observed dust production rates in active
Centaurs.  The outgas rate depends almost exponentially (not as
$1/r^2$) on the illumination-weighted orbital separation, which in
turn depends on the semimajor axis and the eccentricity.  Moreover,
for all but one of the Centaurs observed to be active we predict
${\dot m}_\mathrm{CO} \gtrsim 10^{-8} \g\;\cm^{-2}\s^{-1}$, the
criterion for dust lifting in Eq.~\refnew{eq:dustlift}. Similarly,
most Centaurs observed to be inactive have predicted CO outgas rates
below $10^{-8} \g\;\cm^{-2}\s^{-1}$, though a handful of inactive
objects with high predicted CO outgas rates exist.

For Chariklo we predict a low anuual CO outgas rate, $10^{-9}
\g\;\cm^{-2}\s^{-1}$; typically it should therefore be inactive, as is
observed \citep{Fornasier}.  However, we argue that it should exhibit
short CO outbursts, occurring twice an orbit after equinoxes and
lasting of order a few months or longer. This would lift up the dust
grains and be observed as a flaring in brightness or an excess in CO
emission. Unfortunately, we know of no photometry of Chariklo after
the last equinox in 2008 \citep{duffard14}, and the next one is not
until 2039.  For Chiron, two outbursts were reported in 1989
\citep{Meech} and 2001 \citep{RM,Jewitt}. While \citet{ortiz15}
propose that Chiron was most likely at equinox in 1983 and 1999,
Chariklo's pole position is highly uncertain. Available lightcurves of
Chiron are insufficient to allow a good pole solution
\citep{person16}. Also, although the \citet{ortiz15} analysis is based
on their interpretation of the \citet{ruprecht15} occultation data as
rings, \citet{ruprecht15} themselves favor a jet/outburst
interpretation. Nonetheless, if the \citet{ortiz15} equinox times are
established firmly, they would indeed cast doubt on our model for ring
formation at least for Chiron's case and demand a new explanation for
Chiron's outbursts.

\section{Summary}\label{sec:summary}

Assuming the width variations seen in Chariklo's inner ring are
long-lived, they indicate that the ring is apse-aligned, that its
eccentricity changes by at least several parts in a thousand between
its inner and outer edges, and, by analogy to the Uranian rings,
that its overall eccentricity may be a few percent or more. If the
apse alignment is maintained by a balance between self-gravity within
the ring and Chariklo's large $J_2$ moment, we find the total ring
mass is of order a few$\times 10^{16}$~g. If the ring particles form a
monolayer, they have typical size of order a few meters and velocity
dispersion of order a few mm/s. These figures are a few times smaller
than those observed in Saturn's main rings \citep{zebker85,french00}
and are within the observational bounds for those of the $\epsilon$ ring
of Uranus \citep{french91}.

Our estimates of the ring particle size and velocity dispersion
indicate that the ring's collisional spreading time is of order $10^5$
years, somewhat shorter than the typical Centaur dynamical lifetime of
a few Myr \citep{bailey09}. This short spreading time favors ring
formation scenarios that occur during or after Chariklo's move from
the Kuiper belt to its current location among the giant planets.
Thus, while impacts from $\sim$10~km bodies during Chariklo's few-Gyr
sojourn in the Kuiper belt could easily have thrown enough ejecta into
orbit to form rings, it is unlikely that rings so formed would have
lasted long enough to be the ring system we see today. Perhaps more
likely is ring formation from an initially close-in satellite pushed
inside the Roche radius during the close encounter with Neptune that
turned Chariklo into a Centaur. However, this requires a close match
between the satellite's initial orbit energy and the strength of the
Neptune encounter. That 2060 Chiron likely also hosts rings
\citep{duffard14,ortiz15} suggests that Centaurs with rings may occur
more often than this formation scenario can easily accommodate.

We also consider ring formation via a very different process: dust
particles lifted from Chariklo into close orbit by an outflow of
sublimated CO.  When Chariklo shifted from a more distant orbit to one
with its current $\sim$16~AU periapse distance, the corresponding
increase in the equilibrium temperature and therefore partial pressure
of CO $\sim 1$~km below the surface could have forced dust particles
off the surface and, after mutual collisions, into ring orbits. If
rings are indeed associated with outgassing, we predict that they
should be ubiquitous among large Centaurs but absent among small
comets and very large KBOs (e.g., Pluto). In addition, our model
  predicts that the CO sublimation rate should surge within a few
  months of an equinox crossing. Further observations to refine the
  pole positions of Chariklo, Chiron, and possibly other Centaurs, and
  to monitor their brightness during the several months following
  equinoxes, would provide a very clear test for ring formation via
  outgassing.

Additional high-cadence and/or multi-chord occultation observations
that further constrain the ring surface density profile, the ring
widths, and Chariklo's shape and orientation within the rings will
help us better assess these and other formation mechanisms for the
Chariklo system. Moreover, some of the proposed ring formation
scenarios suggest that many more Centaurs should have rings or small
moons. We hope that future occultation observations of different
Centaurs can test this prediction.

\acknowledgements We thank Maryame El Moutamid, Phil Nicholson, and an
anonymous referee for knowledgable and thought-provoking comments that
led to improvements in this paper. We also acknowledge financial
support by NSERC. MP thanks Bok Tower Gardens for their warm
hospitality during the final stages of writing.

\bibliographystyle{apj}

\end{document}